\documentclass[]{article}
\usepackage[letterpaper, margin=1in]{geometry}
\usepackage{setspace}
\usepackage{graphicx}
\usepackage{lineno}
\usepackage{multirow}
\usepackage{placeins}
\usepackage{float}
\usepackage{caption}
\usepackage{gensymb}
\usepackage{amsmath, amssymb} 
\usepackage{amsthm}
\usepackage{authblk}
\usepackage{natbib}

\newtheorem{theorem}{Theorem}

\newtheorem{proposition}{Proposition}
\theoremstyle{definition}

\title{SMOOTH DENSITY SPATIAL QUANTILE REGRESSION}
\author[1]{Halley Brantley\thanks{hlbrantl@ncsu.edu}}
\author[2]{Montserrat Fuentes}
\author[3]{Joseph Guinness}
\author[4]{Eben Thoma} 
\affil[1]{NC State University}
\affil[2]{Virginia Commonwealth University}
\affil[3]{Cornell University,} 
\affil[4]{U.S. Environmental Protection Agency} 
\date{\vspace{-5ex}}

\begin{document}
\maketitle

\section*{Abstract}
We derive the properties and demonstrate the desirability of a model-based method for estimating the spatially-varying effects of covariates on the quantile function. By modeling the quantile function as a combination of I-spline basis functions and Pareto tail distributions, we allow for flexible parametric modeling of the extremes while preserving non-parametric flexibility in the center of the distribution. We further establish that the model guarantees the desired degree of differentiability in the density function and enables the estimation of non-stationary covariance functions dependent on the predictors. We demonstrate through a simulation study that the proposed method produces more efficient estimates of the effects of predictors than other methods, particularly in distributions with heavy tails. To illustrate the utility of the model we apply it to measurements of benzene collected around an oil refinery to determine the effect of an emission source within the refinery on the distribution of the fence line measurements. 

\onehalfspacing
\section{Introduction}
Quantile regression offers an important alternative to traditional mean regression for problems where the interest lies not in the center of the distribution but in some other aspect. Since the first quantile regression paper was published by \cite{koenker1978regression}, an immense body of literature has been developed and is reviewed in \cite{koenker2005quantile}. \cite{yu2001bayesian} proposed a form of Bayesian quantile regression employing the Asymmetric Laplace Distribution (ASL) as the working likelihood, due to its similarity to the check loss function used by \cite{koenker1978regression}. Both of these approaches perform separate analyses for each quantile level of interest. When quantiles are estimated separately, there is no guarantee of a valid non-decreasing quantile function. There are several approaches to address this issue. The first one is a two-stage method: in the first-stage the quantiles are fit separately using one of the above methods, and in the second stage the estimates are smoothed to ensure monotonicity. This approach has been taken by a variety of authors including \cite{neocleous2008monotonicity},  \cite{rodrigues2016regression}, and \cite{reich2012bayesian} who used it as a more computationally efficient Bayesian spatial method. \cite{bondell2010noncrossing} embed a constraint that ensures monotonicity into the minimization problem, while \cite{cai2015estimation} use prior specifications to ensure constraints in the Bayesian framework. 

The final approach, which we will adopt and extend, is to model the entire quantile function jointly using basis functions. This is the approach taken by \cite{reich2012bayesian} and others \citep{reich2012spatiotemporal, smith2015multilevel} and is more naturally implemented using a Bayesian framework. Regardless of the approach taken, ensuring monotonicity requires either some form of distributional assumption, or constraints on the quantile regression coefficients and the parameter space of the predictors. \cite{cai2015estimation} demonstrated that when predictors are constrained to be positive, the quantile function is monotonic for every possible predictor value if and only if the basis functions are monotonic. This is the approach taken by Zhou et al. (\citeyear{zhou2011calibration, zhou2012}) who first proposed the I-spline quantile regression model whose properties we derive in this paper. 

As in mean regression, a method of incorporating spatial correlation into quantile regression is to model spatially-varying parameters using Gaussian process priors.  \cite{lum2012spatial} use the ASL for the likelihood and incorporate spatial correlation by modeling the error as a function of a Gaussian process and an independent and identically distributed exponential random variable. For large datasets they propose an asymmetric Laplace predictive process, extending the method introduced by \cite{banerjee2008gaussian}. However, the use of the ASL does not allow for valid posterior inference because it does not represent the true likelihood of the observations. \cite{yang2015quantile} combined spatial priors with their Bayesian empirical likelihood approach for modeling the conditional quantiles in the presence of both predictors and spatial correlation, but their method only allows for effects to be estimated at a small fixed number of quantile levels. Several previous methods of modeling a spatially varying conditional quantile function using basis functions have also been advanced \citep{reich2012bayesian, reich2012spatiotemporal}.

We consider the model first proposed by Zhou et al. (\citeyear{zhou2011calibration, zhou2012}) where the quantile function is modeled as a combination of I-splines and the Generalized Pareto Distribution (GPD). The GPD is used to model the tails because it has been shown to be the natural choice for exceedances over a threshold \citep{davison1990models} and provides flexibility as a result of the shape parameter which controls boundedness and the existence of moments. A full description of the I-spline quantile regression model for both independent and spatially correlated data is given in Section 2. In this paper, we formulate the conditions under which the resulting density has the desired degree of differentiability and derive the marginal expectations and spatial covariances which can be non-stationary (Section 3). Our simulation studies demonstrate that ensuring a smooth density can lead to more accurate effect estimates and predictive distributions, compared with methods that do not ensure differentiability (Section 4). We apply the method to benzene measurements from a petrochemical facility to determine the effects of emission sources on concentrations (Section 5). 

\section{Proposed Model}

We model the quantile function of the stochastic process $Y(s)$ as a linear combination of the predictors:

\begin{equation}\label{2.1}
Q(\tau|s, \mathbf{x}(s))  = \beta_0(\tau, s) + \sum_{p=1}^P x_p(s)\beta_p(\tau, s),
\end{equation}
where $\mathbf{x}(s) = (x_1(s), ..., x_p(s)) \in \mathbb{R}^p_+$, is the vector of predictors observed at location $s$, $\beta_0(\tau, s)$ is the quantile function at location $s$ when all predictors are 0, and $\beta_p(\tau, s)$ is the effect of predictor $p$ on quantile level $\tau$ at location $s$. We further follow the approach of \cite{zhou2011calibration} and model  $\beta(\tau, s)$ as a linear combination of I-spline basis functions in the center of the distribution.  We denote the $m^{th}$ I-spline basis function evaluated at $\tau$ as $I_m(\tau)$ and define the constant basis function $I_0(\tau)=1$ for all $\tau$. While I-splines allow for a large degree of flexibility in the center of the distribution, unbounded distributions cannot be estimated using I-splines with a finite number of knots. To solve this issue we use the quantile function of the GPD to model the relationship of the covariate(s) to the process in the tails of the distribution. The model for $\beta_p(\tau, s)$ can then be expressed as

\begin{align}\label{2.2}
\beta_p(\tau, s) & = \begin{cases}
\theta_{0,p}(s) - \frac{\sigma_{L,p}(s)}{\alpha_{L}(s)}\left[\left(\frac{\tau}{\tau_L}\right)^{-\alpha_{L}(s)} - 1 \right]  & \tau < \tau_L  \\
\sum_{m=0}^M\theta_{m,p}(s)I_m(\tau) & \tau_L \le \tau \le \tau_U \\
\left[\sum_{m=0}^M\theta_{m,p}(s)\right] + 
\frac{\sigma_{U,p}(s)} {\alpha_{U}(s)}\left[\left(\frac{1-\tau}{1-\tau_U}\right)^{-\alpha_{U}(s)} - 1 \right]& \tau > \tau_U, \\
\end{cases}
\end{align}
where $\tau_L$ and $\tau_U$ are the thresholds between the tails and the center of the distribution, $\theta_{0,p}$ is the location parameter at the lower tail, and $\theta_{m,p}(s)$ represents the coefficient of the $m^{th}$ I-spline basis function and $p^{th}$ predictor at location $s$. I-splines are monotonic polynomials formed by integrating normalized B-splines (Fig. 1) \citep{ramsay1988}. They are defined on a sequence of knots $\{\tau_0 =...=\tau_{k} < ... < \tau_{M+1} =...=\tau_{M+1+k}\}$, where $k$ represents the degree of the polynomial and $M$ is the number of non-constant basis functions. 

The GPD has three parameters: the shape parameter $\alpha$, the scale parameter $\sigma$, and a location parameter $\mu$. In our parameterization, the location parameter of the lower tail is equal to $\theta_{0,p}(s)$ and the location parameter of the upper tail is equal to $\sum_{m=0}^M\theta_{m,p}(s)$ to ensure the quantile function is continuous. We denote the shape parameters of the lower and upper tails as $\alpha_L(s)$ and $\alpha_U(s)$, respectively, and the scale parameters as $\sigma_{L,p}(s)$ and $\sigma_{U,p}(s)$. We require the shape parameter to be constant across predictors in order to ensure that the density in the tails follows a parametric distribution.  The scale parameters vary by both predictor and location and allow the predictors to affect the tails differently. When $\alpha < 0$, the support of GPD is also bounded above, otherwise the domain is unbounded above. The case when $\alpha = 0$ is interpreted as the limit when $\alpha \rightarrow 0$, i.e.  $\frac{\sigma_{U,p}}{\alpha_{U}} \left[\left(\frac{1-\tau}{1-\tau_U}\right)^{-\alpha_{U}} - 1 \right]$ is replaced with $-\sigma_{U,p}\log\left(\frac{1-\tau}{1-\tau_U}\right)$. The expectation exists if $\alpha$ is less than 1, and the variance exists if $\alpha$ is less than  $1/2$.  

This model formulation ensures a quantile function that is continuous and differentiable at all but a finite number of points. We can thus exploit the result of Tokdar and Kadane (\citeyear{tokdar2012simultaneous}) who demonstrated that a differentiable and invertible quantile function corresponds with the density 
\begin{equation}
f(y) = \frac{1}{Q'(Q^{-1}(y))}.
\end{equation}

\begin{figure}	
	\includegraphics[width = \linewidth]{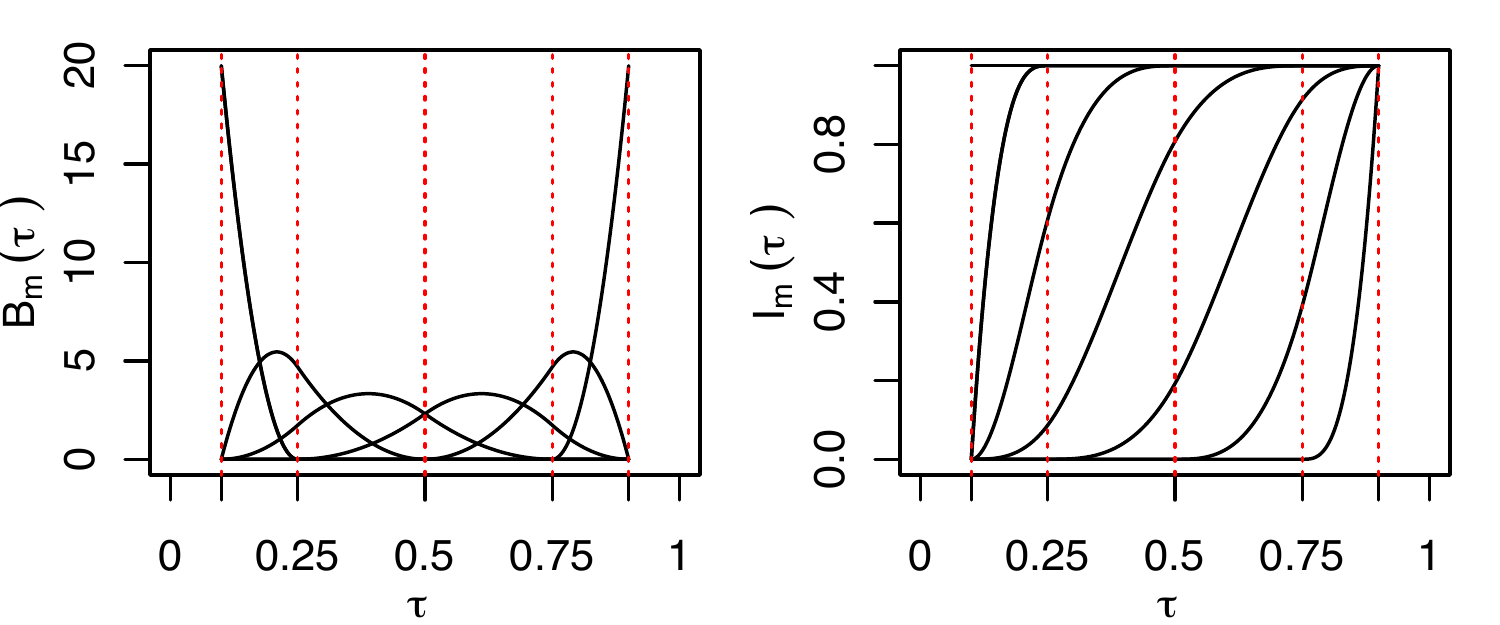}
	\caption{Example set of normalized B-spline (left) and corresponding I-spline (right) basis functions. Dotted vertical lines indicate knot locations.} 
	\label{fig:1}
\end{figure}

To ensure the quantile function is monotonic we introduce latent parameters with Gaussian process priors, $\theta_{m,p}^* \sim \mathcal{GP}(\mu^*_{m,p}, \Sigma^*_{m,p})$ and define $\theta_{0,p}(s) = \theta^*_{0,p}(s)$ and $\theta_{m,p}(s) = \exp{\theta^*_{m,p}(s)}$ for $m>0$. By using this formulation the resulting $\theta_{m,p}(s)$ are modeled as log Gaussian processes. No constraints are placed on $\theta_{0,p}$ which allows predictors to have a negative effect on the response. 

The model formulation has many advantages including the ability to allow the effect of each predictor to vary by quantile level and by spatial location while guaranteeing a valid quantile function. It can also accommodate a variety of tail distributions including both bounded and unbounded tails. Furthermore, we show in Section 3 that we can guarantee the degree of differentiability of the corresponding density function. 

\cite{reich2012spatiotemporal} proposed a similar model, constructing the quantile function using parametric Gaussian basis functions. While the parametric basis functions allow for straightforward evaluation of the density, they do not guarantee a differentiable quantile function, which results in a non-continuous density function (Fig. 2). We show through both simulation and applied data analysis that constraining the density to be continuous and differentiable can result in better parameter estimates and out-of-sample scores.    

\begin{figure}
	\includegraphics[width = \linewidth]{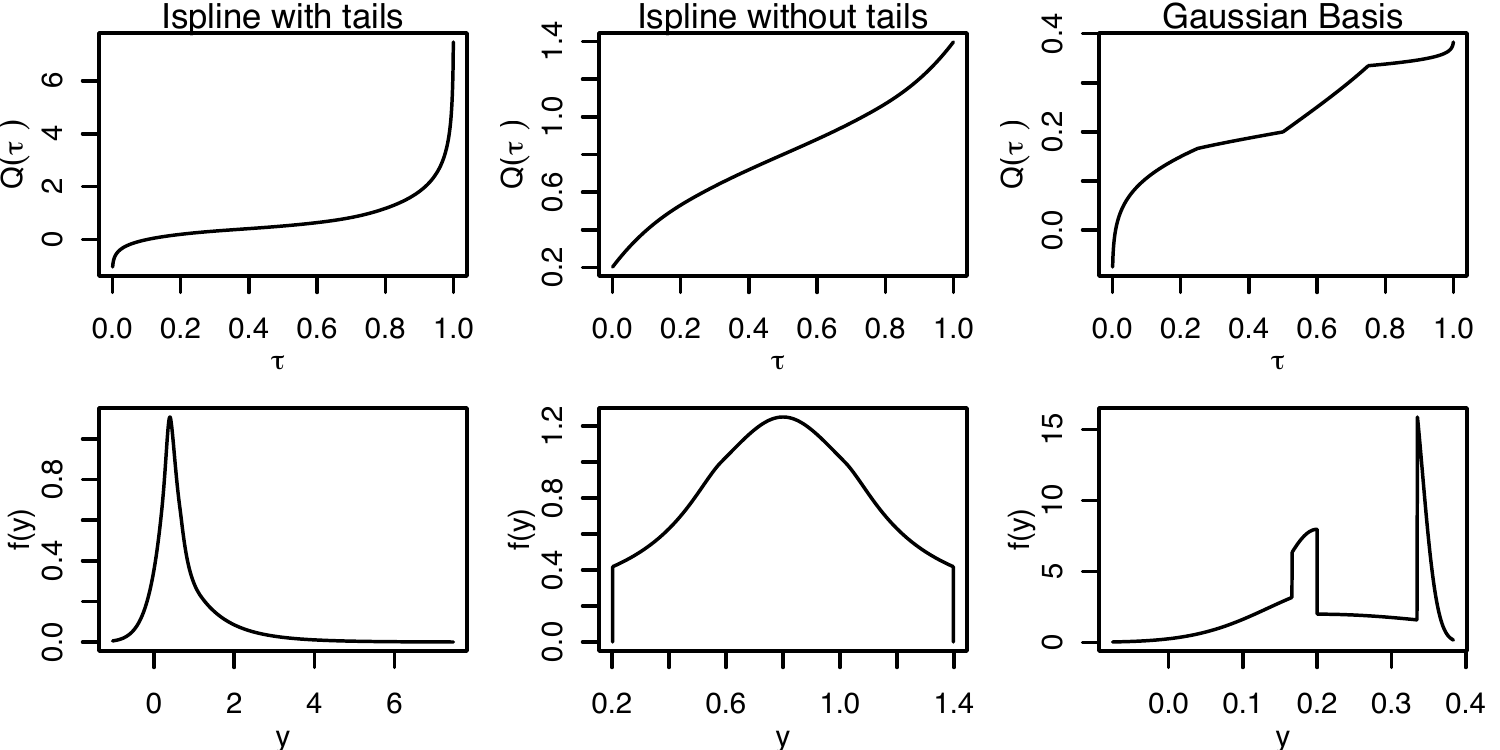}
	\caption{Examples of quantile function (top row) and corresponding density function (bottom row) constructed using different bases. } 
	\label{fig:2}
\end{figure}

\section{Model Properties}
\subsection{Validity Of Quantile Function}
Assuming an I-spline order $k>1$, the proposed quantile function is continuous everywhere and is differentiable for all values of $\tau \in (0,1)$ except $\tau_L$ and $\tau_U$. Thus, a necessary and sufficient constraint to ensure a valid quantile function is $Q'(\tau) \geq 0$ for all $\tau$ at which the derivative exists. For all values of $\tau$ such that $\tau_L < \tau < \tau_U$,  $Q'(\tau) =\sum_{m=1}^M B_m(\tau) \sum_{p=1}^P \theta_{m,p}x_p$. By definition, $B_0(\tau) = 0$ for all $\tau$ and $B_m(\tau) \ge 0$ for all $m$ and $\tau$.  Without loss of generality, we will henceforth assume that the predictors are all non-negative, i.e. $\mathbf{x} \in \mathbb{R}_{+}^P$, therefore a sufficient constraint to ensure a valid quantile function is $\theta_{m,p} \ge 0$ for all $p$ and $m>0$.  If $\sigma_{L,p} > 0$ for any $p$ and $\tau \le \tau_L$, $Q'(\tau) = \sigma_{L,p}x_p\tau^{-\alpha_{L}-1}\tau_L^{\alpha_{L}} > 0$. Similarly if $\sigma_{U,p} > 0$ for any $p$, $Q'(\tau) > 0$ when $\tau \ge \tau_U$.

\subsection {Continuity and Differentiability}

In many cases, such as the application described below, it is desirable to ensure that the density is continuous and smooth. Proposition 1 establishes the conditions for continuity of the density function.

\begin{proposition}
	Let $Y$ have a quantile function as defined in (\ref{2.1}) and (\ref{2.2}) with $\sigma_{L,p} > 0$ for at least one $p$, then the density of $Y$ is continuous at $Q(\tau_L|\mathbf{x},\Theta)$ for any $\mathbf{x} \in \mathbb{R}_{+}^P$  if and only if  
	
	\begin{equation}\label{ContinuityConstraintLower}
	\theta_{1,p}  = \frac{\sigma_{L,p}}{\tau_L I_1'(\tau_L)},
	\end{equation}	
	for all $p$. Similarly, given  $\sigma_{U,p} > 0$ for at least one $p$, the density of $Y$ is continuous at $Q(\tau_U|x,\Theta)$  if and only if   
	\begin{equation}\label{ContinuityConstraintUpper}
	\theta_{M,p}  = \frac{\sigma_{U,p}}{(1-\tau_U)I_M'(\tau_U)}.
	\end{equation}
	
\end{proposition}

Having clarified the conditions for a continuous density, which can be viewed as $0^{th}$ order differentiability, Theorem 1 proceeds to establish the conditions for $q^{th}$ order differentiability of the density function of $Y$. 

\begin{theorem} Let $Y$ have a quantile function as defined in (\ref{2.1}) and (\ref{2.2}) with an I-spline basis order greater than $q+1$ and a density that is continuous and $(q-1)^{th}$ order differentiable at $Q(\tau_L)$. If $\alpha_{L}$ is constrained so that Eq. \ref{differentiableConstraint} does not result in $\theta_{q+1,p} < 0$, then Y has a density that is $q^{th}$-order differentiable at $Q(\tau_L)$ for any $\mathbf{x} \in \mathbb{R}_{+}^P$ if and only if 
	
	\begin{equation}\label{differentiableConstraint}
	\theta_{q+1,p} = \frac{1}{I_{q+1}^{(q+1)}(\tau_L)}\left\{\frac{-\sigma_{L,p}}{\alpha_{L}\tau_L^{q+1}}(-\alpha_{L}-q)_{q+1} -  \sum_{m=1}^{q} \theta_{m,p}I_m^{(q+1)}(\tau_L)\right\}
	\end{equation}
	
	where $I_{q+1}^{(q+1)}(\tau_L)$ is the $(q+1)^{th}$ order derivative of the $(q+1)^{th}$ I-spline basis function, $(-\alpha_{L}-q)_{q+1} = \prod_{j=0}^{q}(-\alpha_L - j)$.
	
\end{theorem}

The conditions which guarantee differentiability at $\tau_U$ are similar and are given in the supplemental information. Combined with the positivity constraint on the $\theta$s, these results imply that the shape parameters have an upper bound that is a function of the knot placement. Ensuring a density that is first order differentiable results in the possible values for $\alpha_{L}$ being bounded above by $-1 -\tau_L\frac{I^{(2)}_1(\tau_L)}{I'_{1}(\tau_L)}$. This bound is a function of $I'_1(\tau_L)$ and $I^{(2)}_1(\tau_L)$ which are functions of the first two knot locations. We can still model any tail behavior provided the outermost knots are placed sufficiently close. 

\subsection{Expectations and Covariance}

While our models allow for flexible non-Gaussian distributions, sometimes the first two moments are of interest (e.g., for best linear unbiased prediction). We now elaborate on the various types of covariance structure that can be estimated using the proposed model. We model the covariances of the latent parameters $\theta_{m,p}^*$ using covariance function $C$ such that $Cov[\theta_{m,p}^*(s), \theta_{m,p}^*(s')] =\eta^{2}_{m,p} C(s,s')$ and $Var[\theta^*_{m,p}(s)] = \eta^{2}_{m,p} + \lambda_{m,p}^2$. Consequently, the expectation of $\theta_{m,p}$ can be expressed as $E[\theta_{m,p}] = \mu_{m,p} = \exp[\mu^*_{m,p} + (\eta^{2}_{m,p} + \lambda_{m,p}^2)/2]$ and the covariance of $\theta_{m,p}$ is
\begin{equation}
\Sigma_{m,p}(s,s') = \mbox{Cov}[\theta_{m,p}(s), \theta_{m,p}(s')] = \mu_{m,p}^2(\exp[\eta^{2}_{m,p} C(s,s')] - 1).
\end{equation}
In this section we describe the covariance of the case when $\tau_L = 0$ and $\tau_U = 1$, we elaborate on other cases in the supplementary material. Under these conditions, the conditional expectation of $Y(s)|\Theta(s),\mathbf{x}(s)$ is
\begin{equation}
E[Y(s)|\Theta(s),\mathbf{x}(s)] = \int_0^1 Q_Y[\tau|\Theta(s), \mathbf{x}(s)]d\tau = \sum_m\sum_p \theta_{m,p}(s)x_{p,t}(s)G_m, 
\end{equation}
where $G_m = \int_{0}^{1} I_m(\tau) d\tau$. We further marginalize over the log Gaussian processes $\theta_{m,p}(s)$, with mean $\mu_{m,p}$ and covariance $\Sigma_{m,p}$, to obtain the expectation and covariance of $Y(s)$ 
\begin{equation}
E[Y_t(s)|\mathbf{x}(s)] = \sum_m\sum_p \mu_{m,p}(s)x_{p,t}(s)G_m. 
\end{equation}
\begin{equation}
\text{Cov}[Y(s),Y(s')|\mathbf{x}(s), \mathbf{x}(s')] = \sum_m\sum_pG_m^2x_{p}(s)x_{p}(s')[\Sigma_{m,p}(s, s')]
\end{equation}
Through this simple case we can see that the covariance is dependent on the values of the predictors in addition to the covariance functions of the latent parameters. This dependence on the predictors can result in non-stationary covariances if $x_p$ vary across space, even if $C(s,s')$ is stationary.

\section{Simulation Study}
Our simulation studies demonstrate the superior efficiency of the proposed I-spline quantile regression method (IQR) using four designs from data generating models that are not in the proposed model class (Table \ref{tab:design}). The designs include cases with both light tails (D1 and D3) and heavy tails (D2 and D4), and with (D3 and D4) and without (D1 and D2) spatial correlation. The designs illustrate the flexibility of the proposed method compared with previously established methods. 

For each design the observed response is indexed as $y_t(s_i)$, where $t \in \{1,...,n\}$ indexes the observations at a given location $s_i$ with $i \in \{1,...,S\}$. The predictor vector $\mathbf{x}_{1,t}$ is generated by sampling from a uniform random variable in D1 and D2. In D3 and D4, $\mathbf{z}_t$ is generated by sampling from a Gaussian processes with mean 0, and exponential covariance with range 1 and $\mathbf{x}_{1,t} = \Phi^{-1}(\mathbf{z}_t)$, where $\Phi^{-1}(\tau)$ is the quantile function of the standard normal. The predictor $\mathbf{x}_{2,t}$ is generated by sampling from a uniform random variable in all designs. The observed response is generated by drawing an independent random uniform variable $u_t(s_i)$ and setting:
\begin{equation}
\label{eq:SimDesign}
y_t(s_i) = \beta_0(u_t(s_i), s_i) + \beta_1(u_t(s_i),s_i)x_{1,t}(s_i) + \beta_2(u_t(s_i), s_i)x_{2,t}(s_i).
\end{equation}
In all designs we assume multiple observations are obtained for each location. For each design we simulate $B=50$ independent datasets. In D1 and D2 we simulate 1000 observations per dataset, assuming all observations are from a single location and thus independent. In D3 and D4 we use $S=16$ locations evenly spaced on a unit square and simulate 100 observations per site for a total of 1600 observations per dataset. For each of the datasets we randomly assign 10\% of the data to be used as validation data for the out-of-sample calculations and use the other 90\% as training data. Computational details including a description of the Markov Chain Monte Carlo (MCMC) algorithm and prior specifications are included in the supplementary material.

\begin{table}[t]
	\caption{True parameter functions by design used in the simulation study. The location is given as $s = (s_1, s_2)$, $\Phi^{-1}(\tau)$ represents the quantile function of the standard normal evaluated at $\tau$ and $Q_{Pareto}$ represents the quantile function of the Pareto distribution with the given parameters.}
	\label{tab:design}
	\renewcommand{\arraystretch}{1}
	\begin{tabular}{l | l |l| l}
		\hline
		& $\beta_0(\tau, s)$ & $\beta_1(\tau, s)$ &
		$\beta_2(\tau, s)$\\ 
		\hline
		D1 	   & $0.1\Phi^{-1}(\tau)$ &  0.3$\tau$ 	& $Q_{Pareto}(\tau, \alpha=-0.2, \mu = 0, \sigma = 0.1)$\\
		D2     & $0.1\Phi^{-1}(\tau)$ & 0.3$\tau$ & $Q_{Pareto}(\tau,  \alpha=0.3, \mu = 0, \sigma = 0.3)$\\
		D3 	   & $(.05 + .2s_1s_2)\Phi^{-1}(\tau)$ & $.3e^{s_2} + 0.2\tau$ & $Q_{Pareto}(\tau, \alpha= -0.1, \mu=0, \sigma=.1 )$\\
		D4 	   & $(.05 + .2s_1s_2)\Phi^{-1}(\tau)$ & $.3e^{s_2} + 0.2\tau$ & $Q_{Pareto}(\tau, \alpha=.4s_1, \mu =0.3, \sigma = 0.4)$\\
		\hline
	\end{tabular}
\end{table}

We compare the estimates from the proposed model (IQR) with those from the model using parametric Gaussian basis functions (GAUS) proposed by \cite{reich2012spatiotemporal} and the non-crossing quantile regression estimates (NCQR) proposed by  \cite{bondell2010noncrossing}. For the IQR and GAUS methods the estimates of $\beta(\tau, s)$ represent the means of the corresponding posterior samples. For the NCQR method the estimates of $\beta(\tau, s)$ are obtained by minimizing the check loss function combined with the non-crossing constraint. The GAUS model allows for spatially varying coefficients and spatial correlation while the NCQR method assumes independent and identically distributed samples.  

We index the quantile levels at which the methods are compared by $j \in {1, ..., J}$.  For each quantile level, $\tau_j$, and simulated dataset replicate, $b \in \{1, ..., B\}$, the estimated coefficients $\widehat{\beta_p}(\tau_j, s_i)$, were compared using root mean integrated square error (RMISE). The RMISE for simulated dataset $b$ was calculated for a given $\beta_p$ and sequence $\tau_1, ..., \tau_J$:

\begin{equation}
RMISE(\beta_p)^{(b)} = \sqrt{\frac{1}{S}\sum_{i=1}^S\sum_{j=1}^J\delta_{j}\left[\widehat{\beta_p}(\tau_j,s_i)^{(b)} - \beta_p(\tau_j, s_i)\right]^2}
\end{equation} 
where $\delta_{j} = \tau_j - \tau_{j-1}$. The means and standard errors of the RMISEs as well as the coverage of the 95\% confidence (NCQR) or credible (IQR and GAUS) intervals were then calculated for each method and design (Table \ref{tab:beta_mid}).  

Both IQR and the GAUS method produce density estimates. The NCQR method does not estimate the entire quantile function and therefore can not be used to create a density estimate without substantial additional calculation. To evaluate the predictive densities we use the log score, which is the logarithm of the predicted density evaluated at the training and validation data. This is a strictly proper scoring rule \citep{gneiting2007strictly}. We calculate the log score for each observation as the log of the posterior mean of the predictive density evaluated at the observation. The total log score for each dataset is calculated as the mean of the log scores for the individual observations. The mean and standard error by simulation design are calculated using the total log score values of the 50 simulated datasets.  

\begin{table}[!t]
	\caption{Comparison of fitted $\beta(\tau)$ functions $\tau = (.05, .06, ..., .94, .95)$. COV represents the coverage of the 95\% credible (IQR and GAUS) or confidence interval (NCQR).}
	\label{tab:beta_mid}
	\renewcommand{\arraystretch}{1} 
	\centerline{\tabcolsep=3truept\begin{tabular}{lrrrcrrrcrrr} \hline 
			\multicolumn{1}{l}{\bfseries }&\multicolumn{3}{c}{\bfseries $\beta_0$}&\multicolumn{1}{c}{\bfseries }&\multicolumn{3}{c}{\bfseries $\beta_1$}&\multicolumn{1}{c}{\bfseries }&\multicolumn{3}{c}{\bfseries $\beta_2$}\tabularnewline
			\cline{2-4} \cline{6-8} \cline{10-12}
			\multicolumn{1}{l}{}&\multicolumn{1}{c}{RMISE}&\multicolumn{1}{c}{SE}&\multicolumn{1}{c}{COV}&\multicolumn{1}{c}{}&\multicolumn{1}{c}{RMISE}&\multicolumn{1}{c}{SE}&\multicolumn{1}{c}{COV}&\multicolumn{1}{c}{}&\multicolumn{1}{c}{RMISE}&\multicolumn{1}{c}{SE}&\multicolumn{1}{c}{COV}\tabularnewline
			\hline
			{\bfseries D1}&&&&&&&&&&&\tabularnewline
			~~IQR&$0.014$&$0.001$&$0.92$&&$0.027$&$0.001$&$0.89$&&$0.022$&$0.002$&$0.91$\tabularnewline
			~~GAUS&$0.016$&$0.001$&$0.92$&&$0.022$&$0.002$&$0.93$&&$0.025$&$0.002$&$0.93$\tabularnewline
			~~NCQR&$0.017$&$0.001$&$0.96$&&$0.025$&$0.001$&$0.97$&&$0.026$&$0.001$&$0.97$\tabularnewline
			\hline
			{\bfseries D2}&&&&&&&&&&&\tabularnewline
			~~IQR&$0.019$&$0.001$&$0.93$&&$0.035$&$0.002$&$0.91$&&$0.047$&$0.002$&$0.90$\tabularnewline
			~~GAUS&$0.038$&$0.005$&$0.83$&&$0.065$&$0.009$&$0.83$&&$0.113$&$0.007$&$0.76$\tabularnewline
			~~NCQR&$0.025$&$0.001$&$0.98$&&$0.045$&$0.002$&$0.97$&&$0.051$&$0.002$&$0.97$\tabularnewline
			\hline
			{\bfseries D3}&&&&&&&&&&&\tabularnewline
			~~IQR&$0.029$&$0.001$&$0.95$&&$0.050$&$0.002$&$0.95$&&$0.027$&$0.001$&$0.99$\tabularnewline
			~~GAUS&$0.027$&$0.001$&$0.97$&&$0.046$&$0.001$&$0.97$&&$0.032$&$0.001$&$0.98$\tabularnewline
			~~NCQR&$0.050$&$0.000$&$0.64$&&$0.201$&$0.001$&$0.16$&&$0.026$&$0.002$&$0.92$\tabularnewline
			\hline
			{\bfseries D4}&&&&&&&&&&&\tabularnewline
			~~IQR&$0.038$&$0.001$&$0.94$&&$0.062$&$0.002$&$0.96$&&$0.094$&$0.004$&$0.94$\tabularnewline
			~~GAUS&$0.094$&$0.047$&$0.94$&&$0.104$&$0.034$&$0.95$&&$0.182$&$0.027$&$0.93$\tabularnewline
			~~NCQR&$0.054$&$0.001$&$0.75$&&$0.197$&$0.001$&$0.24$&&$0.112$&$0.002$&$0.84$\tabularnewline
			\hline
	\end{tabular}}
\end{table}

We compare all three methods using $\tau = \{0.05, 0.06, ..., 0.94, 0.95\}$. Four non-constant basis functions per predictor were used in both the IQR and GAUS methods. The results given in Table \ref{tab:beta_mid} demonstrate that while the 3 methods perform similarly for D1 (independent, light tails), the IQR method performs substantially better than GAUS in the heavy-tailed designs (D2 and D4) and substantially better than NCQR in the spatially varying designs (D3 and D4). Compared to the nominal coverage rate of 0.95, the IQR method has good coverage for all of the designs, with the lowest coverage being 0.88 for $\beta_1$ in D1. GAUS had poor coverage for D2, while NCQR had poor coverage for D3 and D4. 

Unlike the NCQR method, both our method and the GAUS method assume parametric forms for the tails and so can be used to estimate parameter effects on extreme quantiles. We compare the parameter estimates for these two methods evaluated at $\tau = \{0.950, 0.951, ..., 0.994, 0.995\}$ in Table \ref{tab:beta_tail}. Our method performs better in all cases except D1 $\beta_1$, which is a linear function of $\tau$. 

\begin{table}[!t]
	\renewcommand{\arraystretch}{1} 
	\caption{Comparison of fitted $\beta(\tau)$ functions 
		$\tau = (.950, .951, ..., .994, .995)$. COV represents the coverage of the 95\% credible (IQR and GAUS) or confidence interval (NCQR).}
	\label{tab:beta_tail}
	\centerline{\tabcolsep=3truept
		\begin{tabular}{lrrrcrrrcrrr}
			\hline\hline
			\multicolumn{1}{l}{\bfseries }&\multicolumn{3}{c}{\bfseries $\beta_0$}&\multicolumn{1}{c}{\bfseries }&\multicolumn{3}{c}{\bfseries $\beta_1$}&\multicolumn{1}{c}{\bfseries }&\multicolumn{3}{c}{\bfseries $\beta_2$}\tabularnewline
			\cline{2-4} \cline{6-8} \cline{10-12}
			\multicolumn{1}{l}{}&\multicolumn{1}{c}{RMISE}&\multicolumn{1}{c}{SE}&\multicolumn{1}{c}{COV}&\multicolumn{1}{c}{}&\multicolumn{1}{c}{RMISE}&\multicolumn{1}{c}{SE}&\multicolumn{1}{c}{COV}&\multicolumn{1}{c}{}&\multicolumn{1}{c}{RMISE}&\multicolumn{1}{c}{SE}&\multicolumn{1}{c}{COV}\tabularnewline
			\hline
			{\bfseries D1}&&&&&&&&&&&\tabularnewline
			~~IQR&$0.0047$&$0.0004$&$0.96$&&$0.0095$&$0.0010$&$0.89$&&$0.0072$&$0.0007$&$0.94$\tabularnewline
			~~GAUS&$0.0051$&$0.0005$&$0.98$&&$0.0089$&$0.0009$&$0.90$&&$0.0077$&$0.0006$&$0.98$\tabularnewline
			\hline
			{\bfseries D2}&&&&&&&&&&&\tabularnewline
			~~IQR&$0.0139$&$0.0014$&$0.95$&&$0.0377$&$0.0022$&$0.85$&&$0.0810$&$0.0051$&$0.76$\tabularnewline
			~~GAUS&$0.0266$&$0.0054$&$0.78$&&$0.0469$&$0.0109$&$0.75$&&$0.0913$&$0.0045$&$0.57$\tabularnewline
			\hline
			{\bfseries D3}&&&&&&&&&&&\tabularnewline
			~~IQR&$0.0094$&$0.0003$&$0.97$&&$0.0124$&$0.0004$&$0.95$&&$0.0089$&$0.0005$&$0.99$\tabularnewline
			~~GAUS&$0.0119$&$0.0004$&$0.95$&&$0.0139$&$0.0005$&$0.99$&&$0.0132$&$0.0005$&$0.99$\tabularnewline
			\hline
			{\bfseries D4}&&&&&&&&&&&\tabularnewline
			~~IQR&$0.0196$&$0.0012$&$0.96$&&$0.0286$&$0.0027$&$0.93$&&$0.1424$&$0.0048$&$0.90$\tabularnewline
			~~GAUS&$0.0802$&$0.0476$&$0.94$&&$0.0666$&$0.0314$&$0.97$&&$0.2007$&$0.0271$&$0.81$\tabularnewline
			\hline
	\end{tabular}}
\end{table}

The results of the log-score comparisons are consistent with the parameter estimates (Table \ref{tab:logScore_comparison}). However, the GAUS method consistently produces higher log-scores in-sample than the IQR method.  Because the likelihood is not constrained to be continuous in the GAUS method, very large likelihood values can be obtained for the in-sample observations (Fig. \ref{fig:2}). In the heavy-tailed designs the IQR method results in higher out-of-sample log-scores.

\begin{table}[!tbhp]
	\caption{Comparison of mean estimated log scores}
	\label{tab:logScore_comparison}
	\renewcommand{\arraystretch}{1} 
	\centerline{\tabcolsep=3truept
		\begin{tabular}{lrrcrr}
			\hline\hline
			\multicolumn{1}{l}{\bfseries }&\multicolumn{2}{c}{\bfseries In-sample}&\multicolumn{1}{c}{\bfseries }&\multicolumn{2}{c}{\bfseries Out-of-sample}\tabularnewline
			\cline{2-3} \cline{5-6}
			\multicolumn{1}{l}{}&\multicolumn{1}{c}{Mean}&\multicolumn{1}{c}{SE}&\multicolumn{1}{c}{}&\multicolumn{1}{c}{Mean}&\multicolumn{1}{c}{SE}\tabularnewline
			\hline
			{\bfseries D1}&&&&&\tabularnewline
			~~IQR&$ 0.339$&$0.003$&&$ 0.315$&$0.008$\tabularnewline
			~~GAUS&$ 0.356$&$0.003$&&$ 0.322$&$0.008$\tabularnewline
			\hline
			{\bfseries D2}&&&&&\tabularnewline
			~~IQR&$-0.223$&$0.004$&&$-0.254$&$0.017$\tabularnewline
			~~GAUS&$-0.219$&$0.005$&&$-0.288$&$0.022$\tabularnewline
			\hline
			{\bfseries D3}&&&&&\tabularnewline
			~~IQR&$ 0.476$&$0.003$&&$ 0.418$&$0.010$\tabularnewline
			~~GAUS&$ 0.536$&$0.003$&&$ 0.419$&$0.009$\tabularnewline
			\hline
			{\bfseries D4}&&&&&\tabularnewline
			~~IQR&$-0.191$&$0.004$&&$-0.238$&$0.012$\tabularnewline
			~~GAUS&$-0.126$&$0.006$&&$-0.287$&$0.022$\tabularnewline
			\hline
	\end{tabular}}
\end{table}

\section{Application}
\subsection{Data}
An amendment to the U.S. National Emission Standards for Hazardous Air Pollutants for petroleum refineries requires the use of two-week time-integrated passive samplers at specified intervals around the facility fence line to establish levels of benzene in the air \citep{fencelinerule}. The utility of fence line measurements as a method of controlling emissions is contingent on their distributions being dependent on nearby sources within the facility. To evaluate the efficacy of passive samplers in monitoring benzene emissions from petroleum refineries, researchers from US EPA Office of Research and Development conducted a year-long field study in collaboration with Flint Hills Resources in Corpus, Christi, TX \citep{thoma2011}. Preliminary analyses found that under consistent wind conditions, downwind concentrations were statistically higher than upwind concentrations \citep{thoma2011}. More sophisticated modeling should be able to shed light on the contributions of individual sources to the concentrations observed at the fence line. Modeling these concentrations requires an extra level of complexity because near-source air pollutant measurements typically exhibit strong spatial correlation along with non-stationary and non-Gaussian distributions even after transformation. Both the spatial covariance and the distribution of the pollutant concentrations can vary as a function of wind and emission source location. Accurately modeling the entire distribution and spatial structure of the pollutant concentrations should improve inference concerning the strengths of known sources. Additionally, due to the stochastic nature of dispersion and variation in background pollutant concentration levels, the effect of a specific source on the pollutant distribution may not be detected through mean regression. Of particular concern both for exposure and compliance evaluation are the source effects on the upper tail of the distribution, in particular the 95\textsuperscript{th} percentile. 

\begin{figure}[h!]
	\centering
	\includegraphics[height = 2.3in]{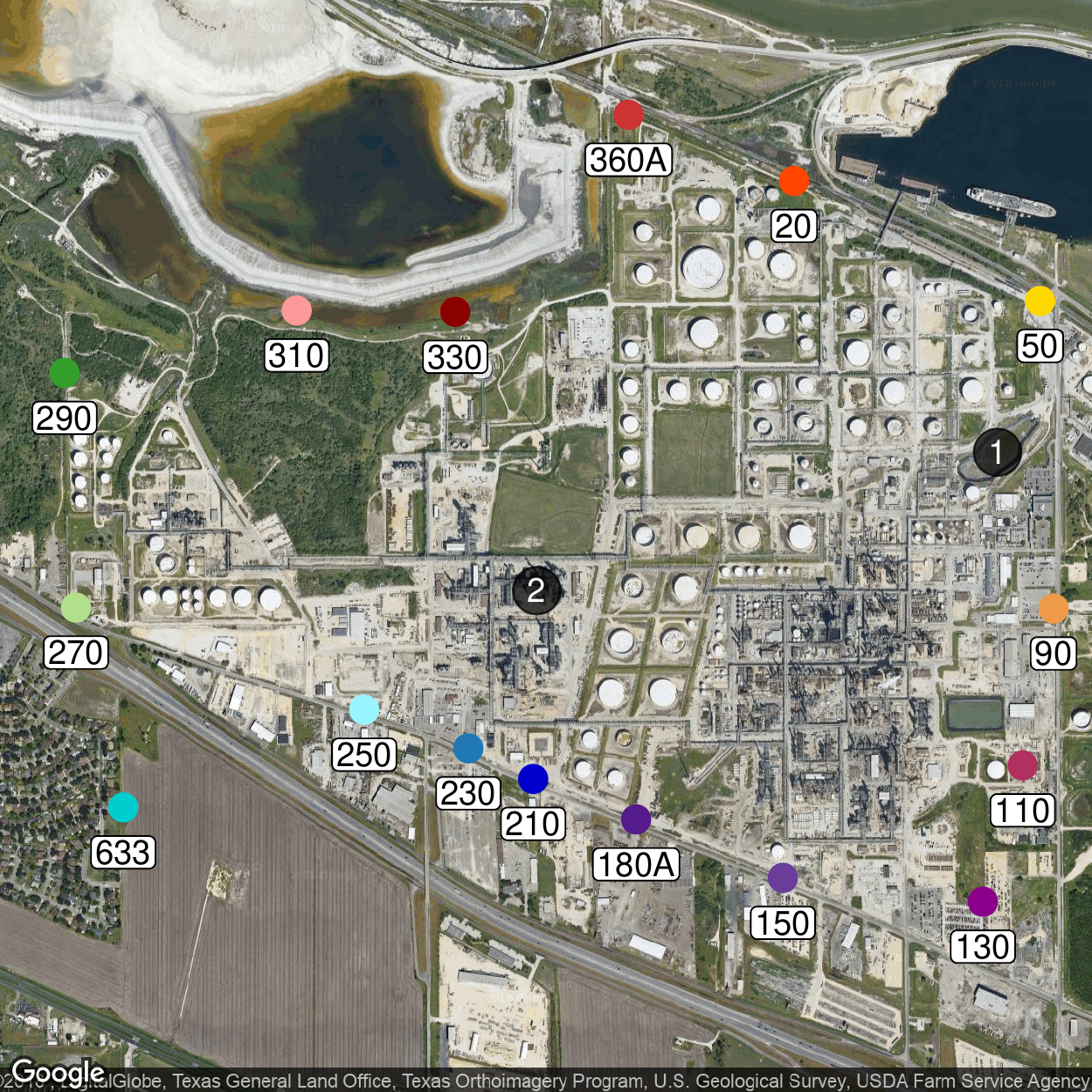}
	\includegraphics[height = 2.3in]{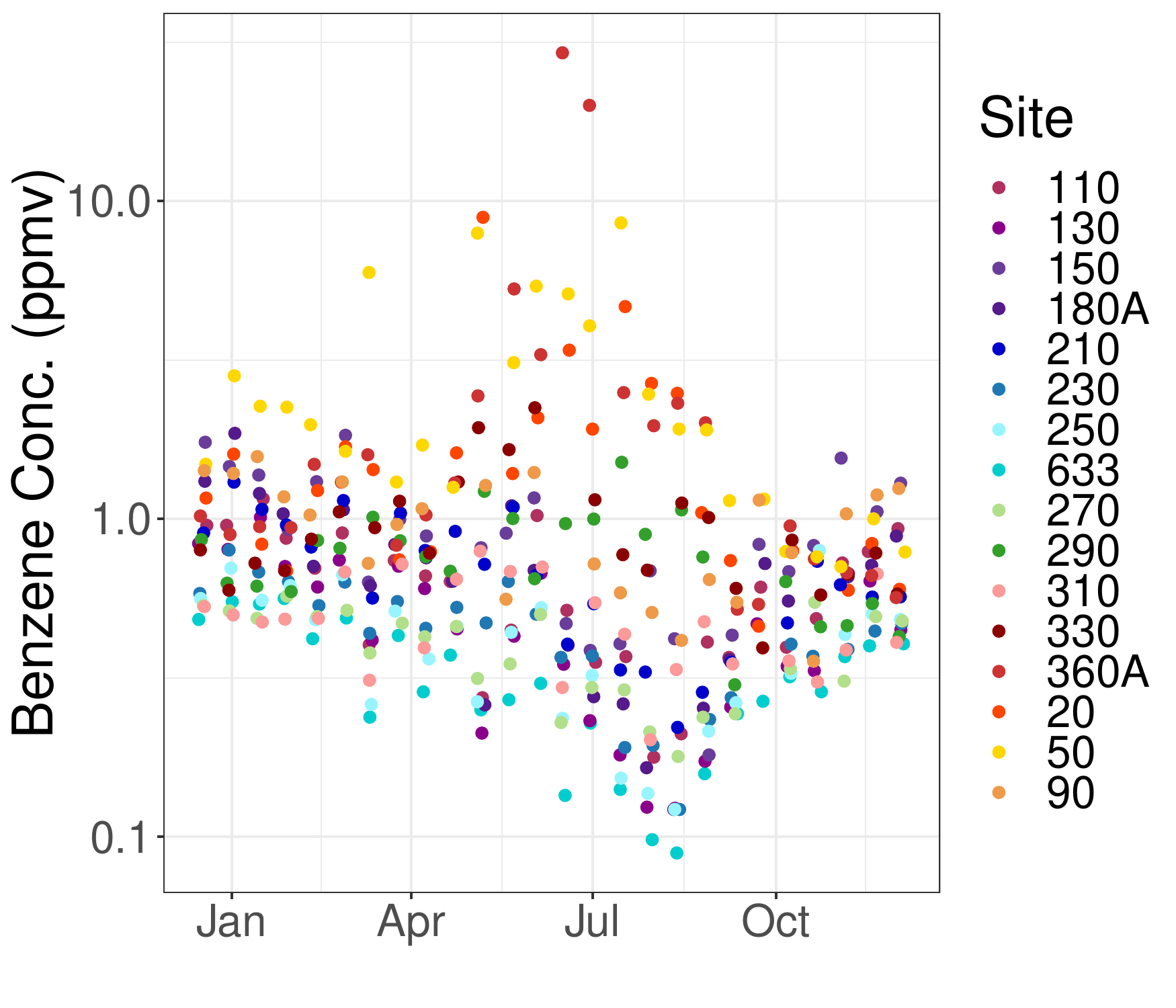}
	\caption{Benzene measurements by time and location. Source locations, $\mathbf{e}_1$ and $\mathbf{e}_2$, are shown in black. Points have been jittered slightly to improve visibility.} 
	\label{fig:flintsummary}
	
\end{figure}

The measurements used in this study were collected between Dec 3, 2008 and Dec 2, 2009 around the Flint Hills West Refinery \citep{thoma2011}. The samplers were attached to the boundary fence around the facility approximately 1.5 m above the ground at 15 locations (Fig. \ref{fig:flintsummary}). In addition, one sampler (633) was deployed at a nearby Texas Commission on Environmental Quality (TCEQ) continuous air monitoring station (CAMS). A total of 406 two-week time-integrated benzene concentration measurements  collected over the course of the year were used in the analysis. Hourly temperature, wind speed and direction were also measured at TCEQ CAMS 633.

\begin{figure}
	\centering
	\includegraphics[width = 2.3in]{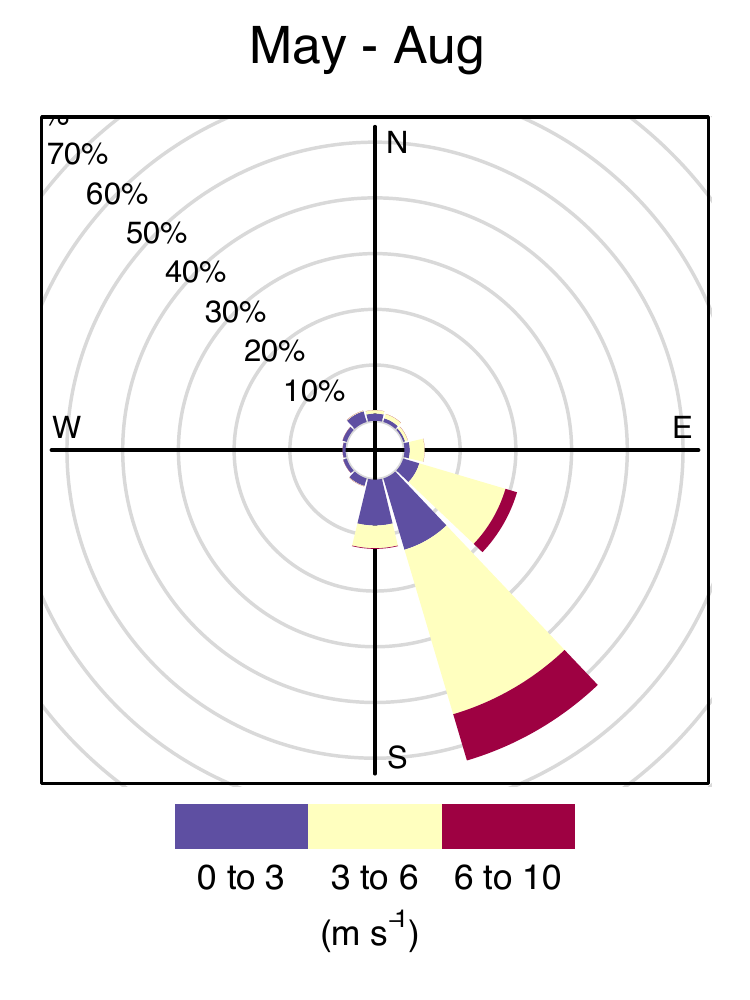}
	\includegraphics[width = 2.3in]{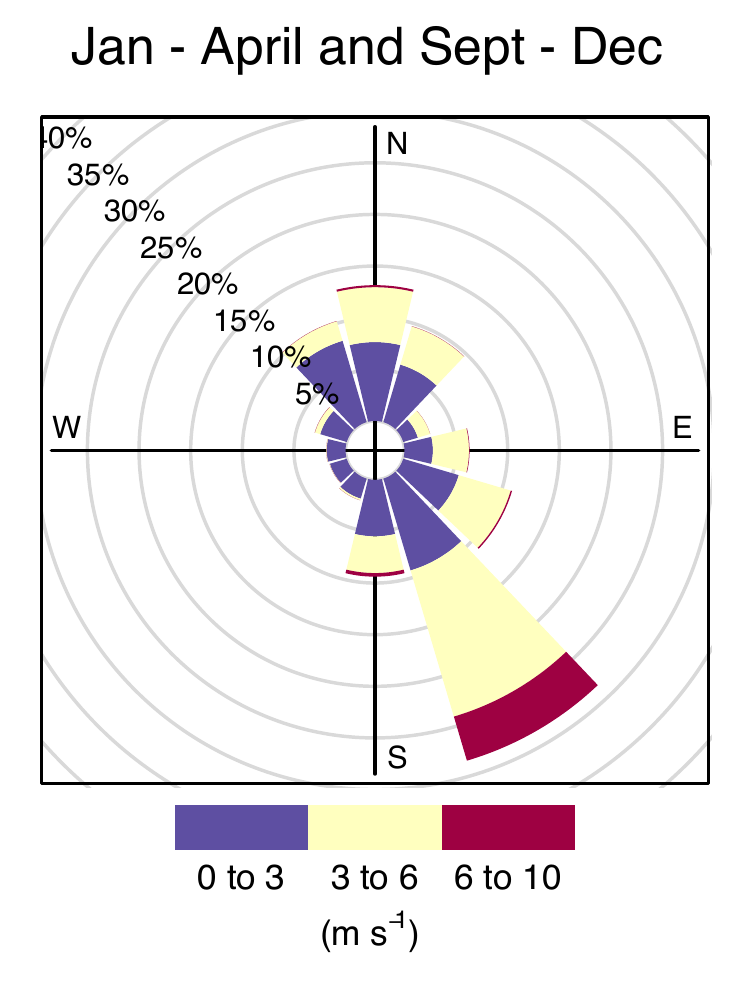}
	\caption{Wind roses for different seasons.}	
	\label{fig:wind1}
\end{figure}

The concentrations exhibited both spatial and temporal trends (Fig. \ref{fig:flintsummary}). In particular, the variance increased dramatically during the summer months. The highest concentrations were observed on the northern edge of the refinery (sites 360, 20, and 50) while the lowest concentrations were observed on the southern edge (sites 250, 633, and 270).  The increase in variance can partly be explained by meteorology (Fig. \ref{fig:wind1}). During the summer the winds consistently blow from the southeast, while during the rest of the year they are more evenly distributed. 

A visual analysis of the concentrations and wind roses for the hourly measurements at each time period suggested that the concentrations were correlated with a source within the refinery. Two probable emission source locations $\mathbf{e}_1$, and $\mathbf{e}_2$ were selected using the reported emission inventory. To determine the effect of the emission sources on the distribution of the benzene concentration, we denote the $t^{th}$ observed value of the benzene concentration at site $s_i$ as $y_t(s_i)$ where $i = 1, ..., 16$ and  $t=1,..., 26$ and will model the quantile function of $Y$ using equation \ref{2.1} and \ref{2.2}. Our full model includes an intercept and three predictors: transport from source 1, transport from source 2, and temperature.   

The predictors that represent transport from a source are calculated from the observed hourly wind vectors and relative spatial locations of the source and measurement. The $t^{th}$ observed value of the transport from source 1 to location $s_i$ is defined as

\begin{equation}\label{app}
x_{1,t}(s_i) = \sum_{h=1}^{336} \left \{ \max \left ( \frac{\mathbf{w}_{t,h} \cdot (\mathbf{e}_1 - \mathbf{s}_i)}{||(\mathbf{e}_1 - \mathbf{s}_i)||}, 0 \right )  \right \}
\end{equation}
where $\mathbf{e}_1$ is the location of emission source 1, and  $\mathbf{s}_i$ is the measurement location. Each hourly wind vector, $\mathbf{w}_{t,h}$, for the two-week period with $h = 1, ..., 336$ was transformed into the same coordinate system and projected onto the vector from the source to the measurement $(\mathbf{e}_1 - \mathbf{s}_i)$. Assuming a constant emission source, the resulting scalar quantity represents the amount of pollutant transported from $\mathbf{e}_1$ to $\mathbf{s}_i$, ignoring the effects of vertical dispersion.  When the wind is blowing from $\mathbf{s}_i$ toward $\mathbf{e}_1$, transport from $\mathbf{e}_1$ will be negative. However, due to finite, small background concentrations, the integrated benzene concentration will remain the same rather than decreasing under these conditions. Therefore the maximum of the transport from $\mathbf{e}_1$ and zero was taken before taking the sum over $h$ in period $t$ (\ref{app}). The transport from source 2 was calculated similarly. 

We use 10-fold cross-validation to determine the most appropriate model for the benzene concentrations. Using each fold as a validation data set, the in-sample and out-of-sample log-score was calculated using both the proposed IQR method and the GAUS method proposed by  \cite{reich2012spatiotemporal} for each combination of predictors (Table 5).  An exponential covariance function and range of 0.5 were used for both methods. The predictors were transformed to be between 0 and 1 before the models were fit. 

For both methods the in-sample log-score tends to increase with the number of predictors included in the model. All of the models with predictors have both higher in-sample and out-of-sample log scores than the intercept only model. Of the models with predictors the one that included all 3 produced the largest out-of-sample log-score for the IQR method, but the lowest out-of-sample log-score for the GAUS method, indicating that adding additional predictors exacerbates the probability of over-fitting by the GAUS method. In all of the cases, our method performs substantially better out-of-sample than the GAUS method.

\begin{table}[t]
	
	\label{tab:logScores}
	\renewcommand{\arraystretch}{1.2}
	\setlength{\tabcolsep}{3pt}
	\caption{Estimated log-scores for training and validation data by method.}
	\vspace{-8pt}
	\begin{center}
		\begin{tabular}{lccccccccccc}
			\hline
			&\multicolumn{5}{c}{In-sample}&&\multicolumn{5}{c}{Out-of-sample}\tabularnewline
			\cline{2-6} \cline{8-12} 
			\multicolumn{1}{l}{Predictors}&\multicolumn{2}{c}{IQR}&&\multicolumn{2}{c}{GAUS}&& \multicolumn{2}{c}{IQR}&&\multicolumn{2}{c}{GAUS}\tabularnewline
			\cline{2-3} \cline{5-6} \cline{8-9} \cline{11-12}
			& Mean & SE && Mean & SE && Mean & SE && Mean & SE\tabularnewline
			None&$-0.29$&$0.01$&&$-0.16$&$0.01$&&$-0.42$&$0.07$&&$-0.45$&$0.09$\tabularnewline
			Source 1&$ 0.04$&$0.01$&&$ 0.49$&$0.01$&&$-0.13$&$0.07$&&$-0.18$&$0.07$\tabularnewline
			Source 2&$ 0.04$&$0.01$&&$ 0.47$&$0.01$&&$-0.14$&$0.08$&&$-0.21$&$0.08$\tabularnewline
			Temperature&$ 0.10$&$0.01$&&$ 0.57$&$0.02$&&$-0.07$&$0.08$&&$-0.25$&$0.08$\tabularnewline
			Source 1 + Source 2&$ 0.21$&$0.01$&&$ 0.81$&$0.02$&&$-0.02$&$0.08$&&$-0.22$&$0.06$\tabularnewline
			Source 1 + Temp&$ 0.30$&$0.01$&&$ 1.00$&$0.03$&&$ 0.08$&$0.08$&&$-0.19$&$0.09$\tabularnewline
			Source 2 + Temp&$ 0.27$&$0.01$&&$ 0.88$&$0.02$&&$ 0.06$&$0.08$&&$-0.24$&$0.09$\tabularnewline
			All&$ 0.39$&$0.01$&&$ 0.95$&$0.04$&&$ 0.13$&$0.09$&&$-0.39$&$0.08$\tabularnewline
			\hline
	\end{tabular}\end{center}
\end{table}

\begin{figure}[h]
	\centering
	\includegraphics[width = 5in]{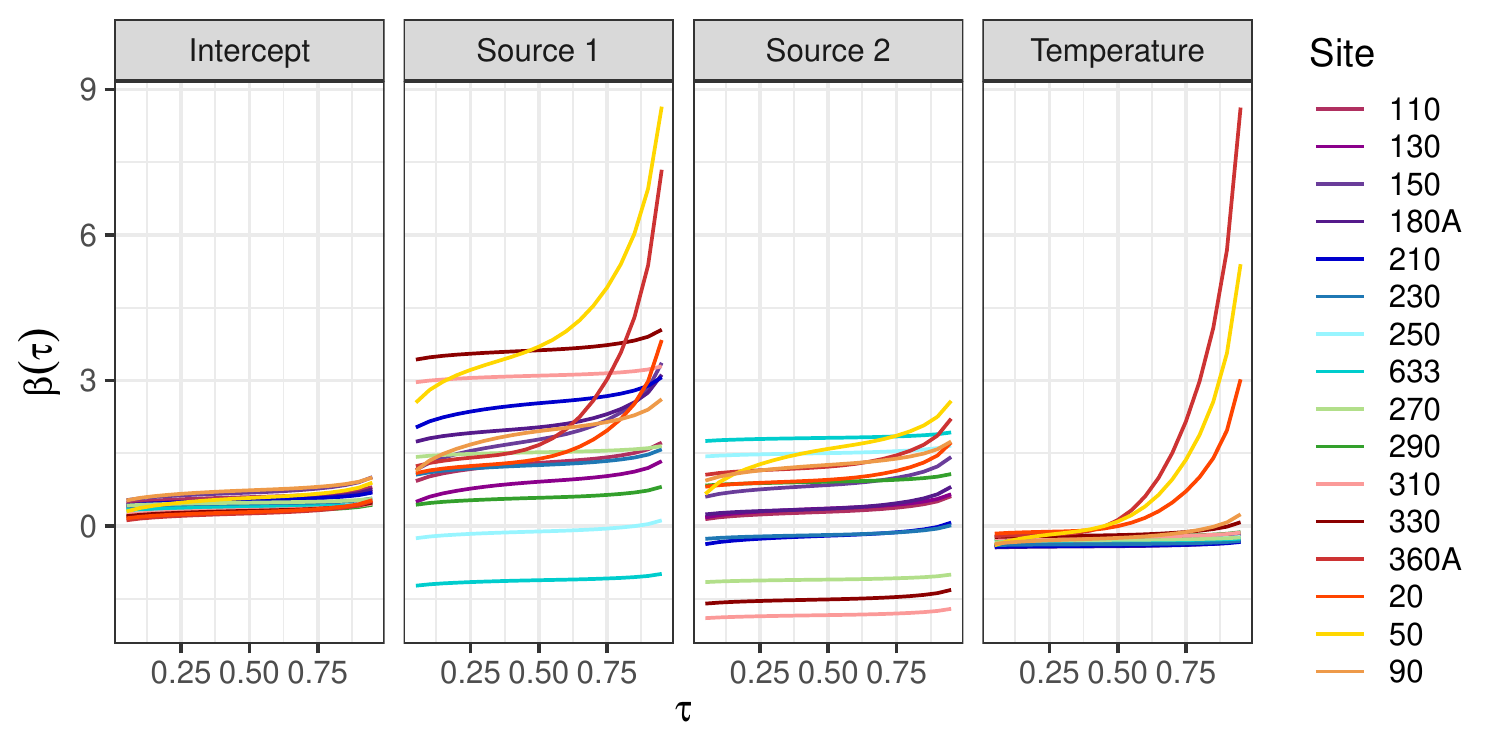}
	\caption{Estimated predictor effect by quantile and location.}	
	\label{fig:coefbytau}
\end{figure}

\begin{figure}
	\centering
	\includegraphics[width = .32\linewidth]{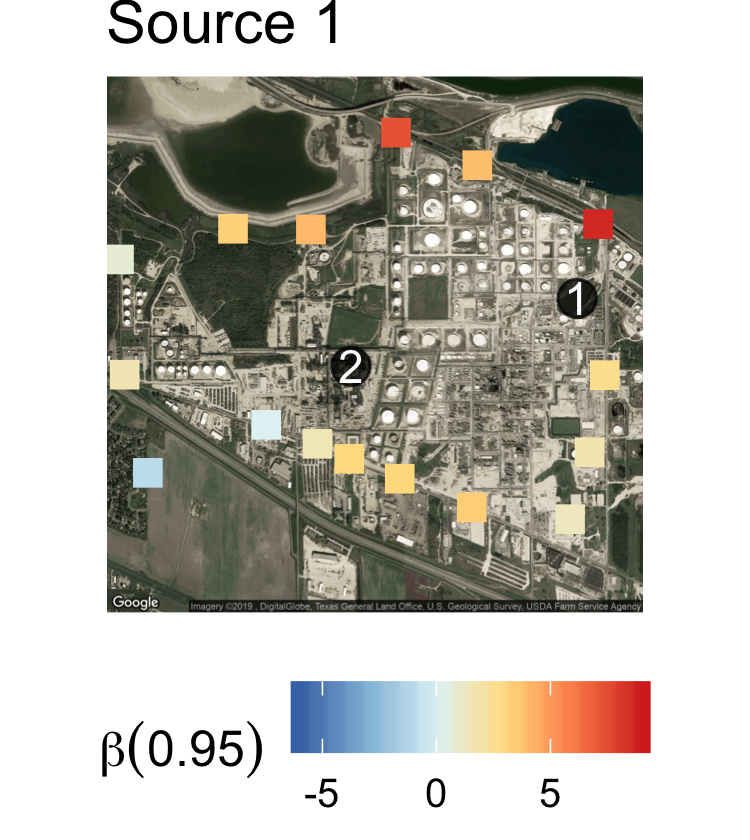}
	\includegraphics[width = .32\linewidth]{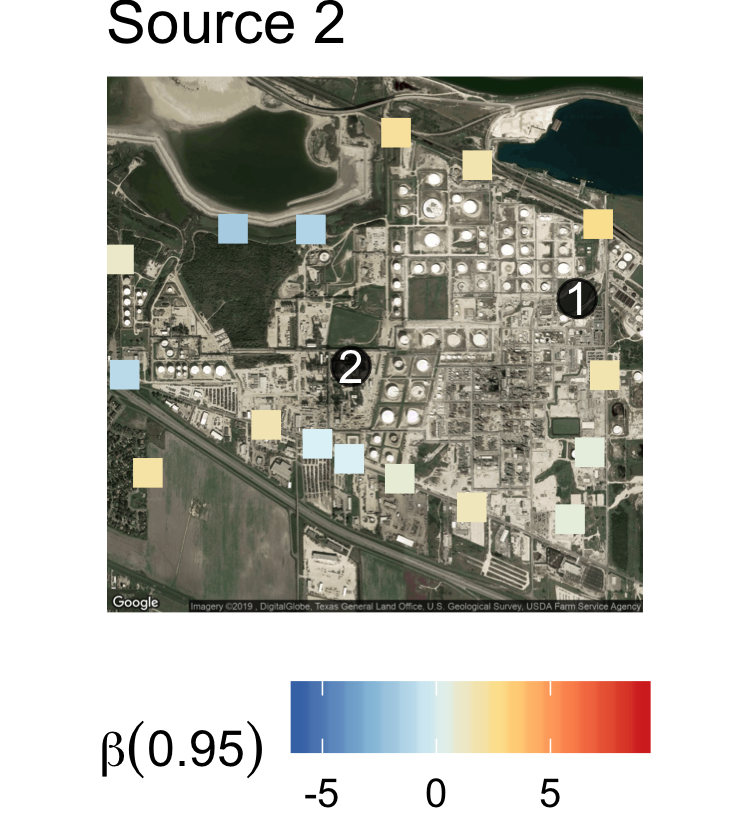}
	\includegraphics[width = .32\linewidth]{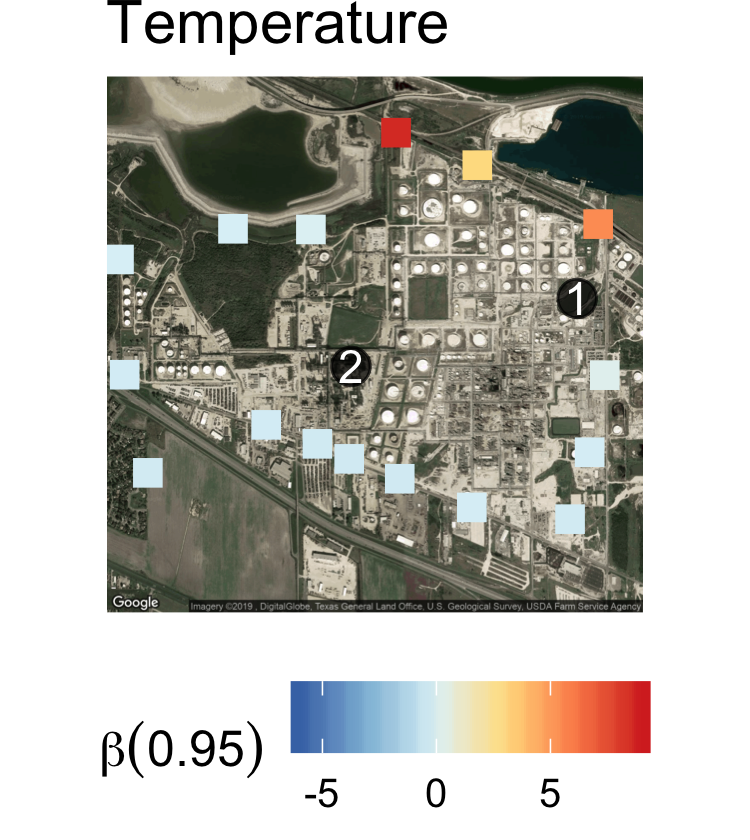}
	\caption{Spatial variation in the effect of the predictors on the $95^{th}$ quantile of fenceline benzene measurements.} 	
	\label{fig:spatialcoef}
\end{figure}

The model was fit to the entire dataset to determine the effects of the sources and temperature on the distribution of benzene at the fence line.  We plot the coefficients by quantile level and location in Fig. \ref{fig:coefbytau}. We can see that the base distribution does not vary as much by location as the effects of the sources and temperature. The effects of the sources on the quantiles of the concentrations range from positive to negative, with the majority of the source effects being positive.  The negative effects could be due to the fact that these sources may not have been constant over the course of the entire year. If wind from a given source corresponded to time points when the source was not emitting it could result in a negative effect on the concentrations.  As can be seen in Fig. \ref{fig:spatialcoef}, the effect of source 1 on the 95\textsuperscript{th} quantile is large and positive for the sites on the northern edge of the refinery and some sites along the southern edge of the refinery. The northern sites were also the locations where the highest concentrations were observed. The effect of source 2 on the 95\textsuperscript{th} quantile was smaller overall and varied by site with positive effects observed on the background site and sites on the northern edge of the refinery (Fig. \ref{fig:spatialcoef}). Temperature also had a strong positive effect on concentrations on the northern edge of the refinery indicating the possibility of another emission source during the summer near the northern edge of the refinery that was not accounted for. 

\section{Discussion}
We have derived the properties and demonstrated the utility of a method for spatial quantile regression that allows for spatially-varying coefficients and flexible tail distributions. By modeling the entire quantile function we exploit the flexibility of non-parametric basis functions in the center of the distribution and the constraints of parametric tails in the areas of the distribution where data is sparse. We have shown the conditions under which the model guarantees a smooth density function with the desired degrees of differentiability and enables the estimation of a non-stationary covariance that is dependent on the predictors. Through both simulations and an application to fence line benzene concentrations we have demonstrated the utility of ensuring a smooth density function with parametric tails and the flexibility and accuracy of the method compared to previous work. 

While the model doesn't currently account for temporal correlation in the response variable, a non-linear function of time could easily be incorporated as a predictor using the current framework. Additionally, temporal correlation could be accounted for by adjusting the priors of the coefficients or incorporating a copula. A multivariate extension for modeling multiple pollutants simultaneously could also be developed through the use of multivariate spatial priors.

\vskip 14pt
\noindent {\large\bf Supplementary Materials}

Proofs of Proposition 1 and Theorem 1 as well as computing details are contained in the supplemental material.  
\par
\vskip 14pt
\noindent {\large\bf Disclaimer}
The views expressed in this publication are those of the authors and do not necessarily represent the views or policies of the U.S. Environmental Protection Agency.

\noindent {\large\bf Acknowledgements}
This work was supported by the National Science Foundation under grant No. 1613219 and the National Institutes of Health under grant No. R01ES027892. The authors would like to thank Oak Ridge Institute for Science and Education for the fellowship funding that supported this work. 
\par

\bibliographystyle{apalike}
\bibliography{references}
\end{document}